\documentclass[prb,
twocolumn,
superscriptaddress,showpacs,amsmath,amssymb]{revtex4}
\usepackage{amsfonts}
\usepackage{bm}
\usepackage{verbatim}

\usepackage{graphicx}

\begin{document}

\title{Polar phase of superfluid $^3$He: Dirac lines in the parameter and momentum spaces}

\author{G.E.~Volovik}
\affiliation{Low Temperature Laboratory, Aalto University,  P.O. Box 15100, FI-00076 Aalto, Finland}
\affiliation{Landau Institute for Theoretical Physics, acad. Semyonov av., 1a, 142432,
Chernogolovka, Russia}

\date{\today}

\begin{abstract}
The time reversal symmetric polar phase of the spin-triplet superfluid $^3$He has two types of Dirac nodal lines. In addition to the Dirac loop in the spectrum of the fermionic Bogoliubov quasiparticles in the momentum space
$(p_x,p_y,p_z)$, the spectrum of bosons (magnons) has Dirac loop in the 3D space of parameters -- 
the components of magnetic field $(H_x,H_y,H_z)$. The bosonic Dirac system lives on the border between the type-I and type-II.
\end{abstract}
\pacs{
}

\maketitle

Originally the topology of the points and lines of level crossing \cite{Born1927,Neumann1929} (diabolical points \cite{Berry1981,Berry2010}) has been investigated in a parameter space. 
In particular, while encircling a diabolical point in the space of two parameters, the wavefunction changes sign.\cite{Berry1981,Berry1984,Berry2010}  Typically this has been applied to electronic spectrum in molecular systems.
Later the topological methods have been applied to the diabolical points in the spectrum of fermionic quasiparticles (Bogoliubov quasipartilces) in gapless superfluids and superconductors,\cite{Volovik1987} where the parameter space is the space of linear momentum in superfluids and quasimomentum in superconductors, or the extended phase space  
$({\bf p},{\bf r})$.\cite{SalomaaVolovik1988,GrinevichVolovik1988,Ryu2010} In particular, the topologically protected diabolical point in 3D momentum space -- the Weyl point -- gives rise to Weyl fermions and effective gauge and gravity fields emerging in the vicinity 
of the Weyl point.\cite{FrogNielBook,Horava2005,Volovik2003} This analog of relativistic quantum field allowed to experimentally verify the Adler-Bell-Jackiw\cite{Adler,BellJackiw}  equation for chiral anomaly  in chiral superfluid $^3$He-A.\cite{Bevan1997}
Then this topological consideration has been extended to the spectrum of bosonic excitations, see e.g.
\cite{Hughes2017,Nakata2017,Balatsky2016,Takahashi2017}.

Recently the new trend is towards the topology in the extended space, which combines the momentum space and the parameter space, see e.g. \cite{Zilberberg2018} 
Here we show that the appropriate system, where the two spaces (momentum space and parameter space) are topologically connected, is the polar phase of superfluid $^3$He discovered in nematically ordered aerogel.\cite{Dmitriev2015}
  
In momentum space the polar phase contains the Dirac nodal line in the quasiparticle spectrum determined by the $2\times 2$ Bogoliubov-Nambu Hamiltonian:
\begin{equation}
{\cal H}(\mathbf{p}) = v_F(p - p_F)\tau^3 + \Delta_P\hat{\bf m}  \cdot \hat{\bf p}  \,\tau^1 \,.
\label{HamiltonianPolar}
\end{equation}
Here $\tau^a$  are the Pauli matrices in the Bogoliubov-Nambu space; $p_F$ and $v_F$ are the Fermi momentum and Fermi velocity in the normal state of liquid $^3$He; $\Delta_P$ is the gap amplitude in the polar phase; $\hat{\bf p}={\bf p}/p$;  $\hat{\bf m}$ is the unit vector of uniaxial anisotropy axis provided by the direction of the aerogel strands, and we choose the coordinate systems with $\hat{\bf z}=\hat{\bf m}$; we ignore here the spin structure of the order parameter  (but later it will be important for the consideration of spin dynamics).

\begin{figure}
\includegraphics[width=\columnwidth]{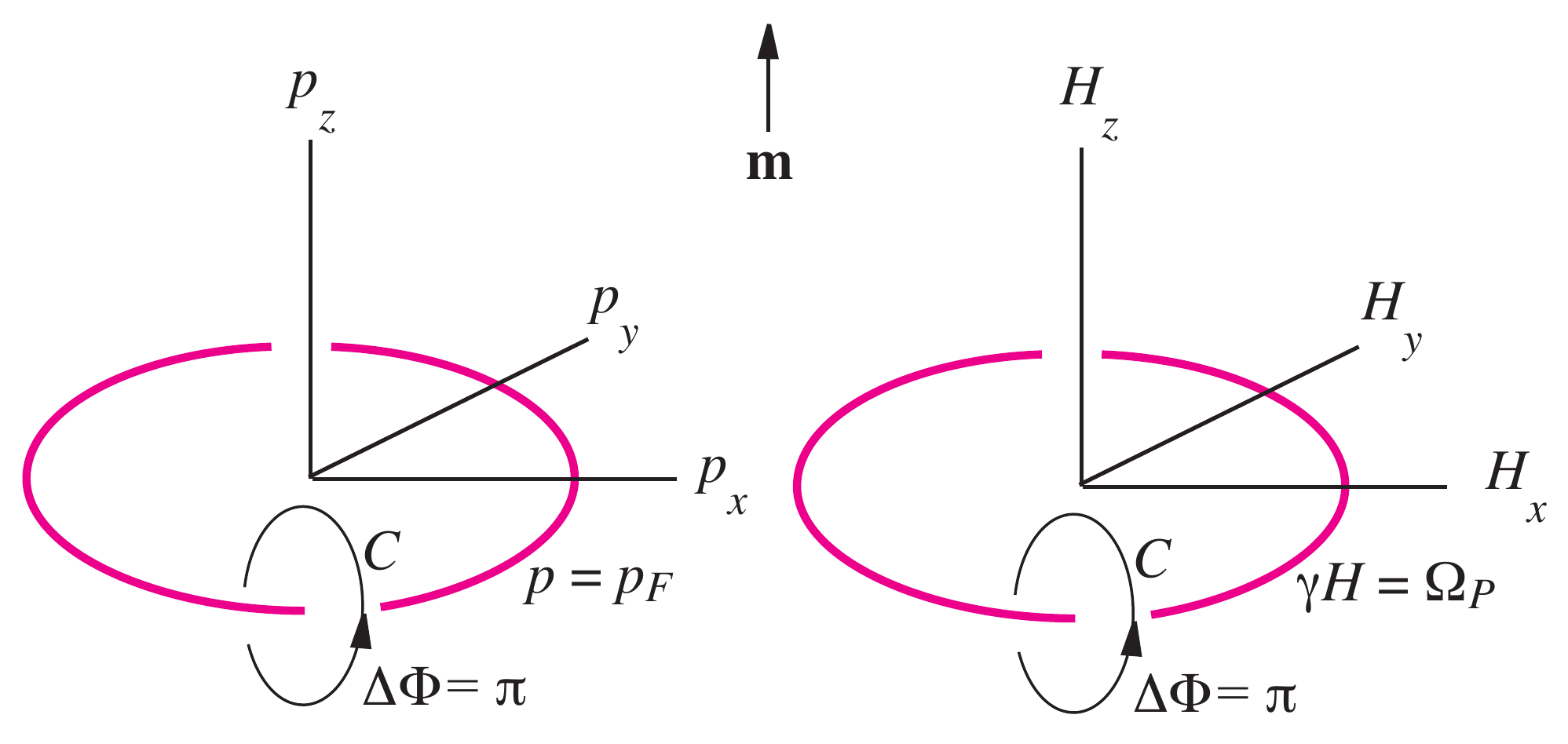}
   \caption{\label{Fig:DiracLine} (Color online)
Exceptional lines of level crossing analyzed by von Neumann and Wigner\cite{Neumann1929} in the polar phase of superfluid $^3$He. The geometric Berry phase around these lines changes by $\pi$.\\
{\it Left}: Dirac line in the quasiparticle spectrum in space of the components of momentum  $(p_x,p_y,p_z)$.
At this topologically protected line  ($p_z=0$, $p=p_F$) the energy of the Bogoliubov quasiparticles in Eq.(\ref{HamiltonianPolar}) is zero.\\
{\it Right}:
Dirac line in the space of parameters -- components of magnetic field $(H_x,H_y,H_z)$, which determine the frequency of magnons in 
Eqs.  (\ref{eq:spectrum1}) and (\ref{eq:spectrum2}). At this topologically protected line ($H_z=0$, $\gamma H=\Omega_P$, where $\Omega_P$ is the Leggett frequency) the branch of optical magnon and the branch of light Higgs mode\cite{Zavjalov2016,VolovikZubkov2015}   cross each other, see Fig. \ref{Fig:Spectrum}.
 }
 \end{figure}

The nodal line, where the spectrum of negative energy states touches  the spectrum of positive energy states,
is  at $p_z=0$ and $p=p_F$, see Fig.\ref{Fig:DiracLine} ({\it left}).  In the vicinity of the Dirac line there emerges the peculiar type of quantum electrodynamics with the non-analytic action for the effective electromagnetic field, $(B^2-E^2)^{3/4}$.\cite{NissinenVolovik2018}

Here we show that the  spectrum of spin waves (magnons) -- the Goldstone modes of  the polar phase -- also experiences the topologically protected Dirac nodal line, but now in the parameter space, see Fig.\ref{Fig:DiracLine} ({\it right}). This spectrum at different magnitudes and orientations of magnetic field has been measured in Ref. \cite{Dmitriev2016}.
The equation  for magnetization ${\bf M}$ in the spin-wave modes follows from the Leggett equation, obtained
using the free energy for spin dynamics, see Ref. \cite{Zavjalov}:
\begin{equation}
F= \frac{1}{2} \chi^{-1}_{ab} M_a M_b - {\bf M}\cdot {\bf H} +
\frac{\chi_\perp}{2\gamma^2}  \Omega_P^2(\hat {\bf d}\cdot \hat {\bf m})^2 \,.
\label{FreeEnergy}
\end{equation}  
Here ${\bf H}$ is the external magnetic field; $\gamma$ is gyromagnetic ratio; the unit vector $\hat {\bf d}$ is the spin part of the order parameter, which determines the easy axis of spontaneous anisotropy of spin susceptibility $\chi_{ab}$. The last term in Eq.(\ref{FreeEnergy}) is the spin-orbit coupling, where $\Omega_P$ is the so-called Leggett frequency, the frequency of the longitudinal NMR.
The equation for magnetization  has the following matrix form:\cite{Dmitriev2016}
\begin{eqnarray}
\label{eq:spectrum1}
\omega^2 \Psi=  {\cal H}(\mathbf{H})\Psi\,,
\\
  {\cal H}(\mathbf{H})=\frac{(\gamma H)^2 +\Omega_P^2}{2} +
\nonumber
\\
 +\left(\frac{(\gamma H)^2 - \Omega_P^2}{2}  +\Omega_P^2  \cos^2\lambda \right)\tau^3
 -  \Omega_P^2 \sin\lambda \cos\lambda \,\tau^1.
\label{eq:spectrum2}
\end{eqnarray}
Here the two-component function is $\Psi=(M_\perp, M_\parallel - M$), where $M_\perp$ and $M_\parallel$ are the transverse and longitudinal components of magnetization with respect to the direction of magnetic field, and $M=\chi_\perp H$ is an equilibrium magnetization; $\tau^a$  are the Pauli matrices connecting the two components of magnetization;    $\omega_L=\gamma H$ is Larmor frequency; $\lambda$ is the angle of magnetic field with respect to anisotropy axis $\hat{\bf m}$, i.e. $\cos\lambda=\hat{\bf m}\cdot {\bf H}/H$.

\begin{figure}
\includegraphics[width=\columnwidth]{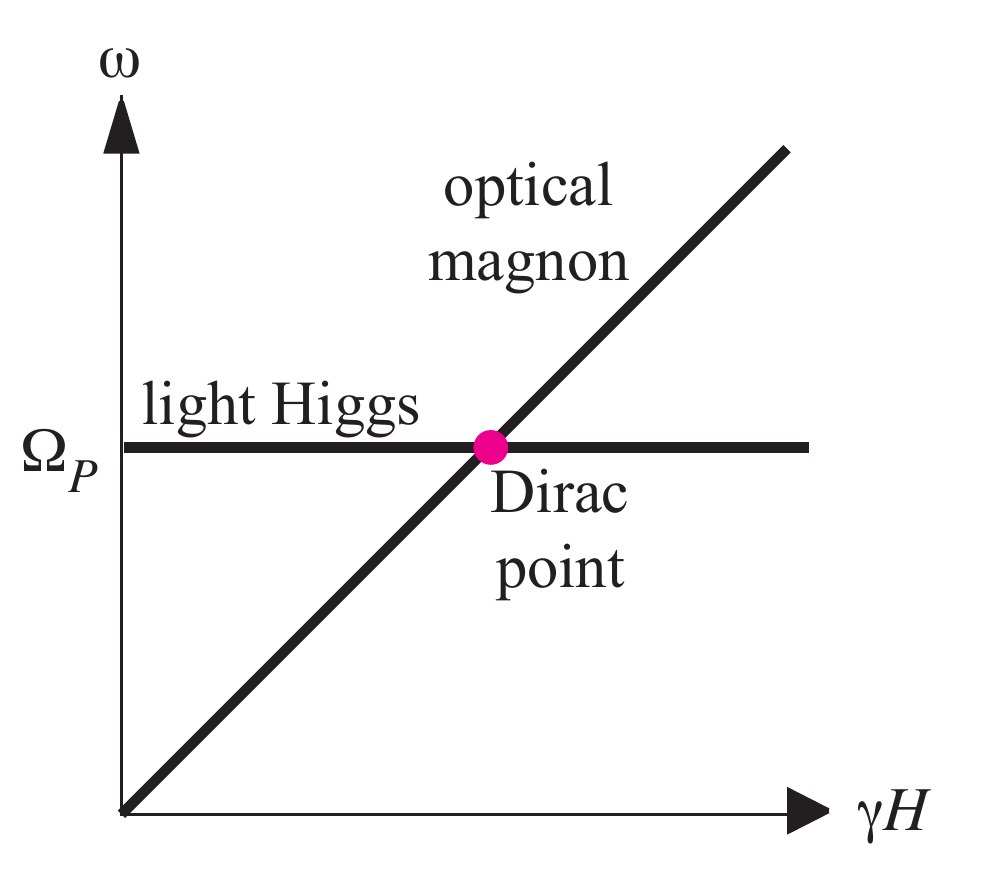}
\caption{\label{Fig:Spectrum} 
 For magnetic field ${\bf H}\perp \hat{\bf m}$, two banches of magnon spectrum (light Higgs mode $\omega=\Omega_P$ and optical magnon $\omega=\gamma H$) do not interact with each other and cross each other at the exceptional point $\gamma H=\Omega_P$. In the 3D space of magnetic field this Dirac point becomes the Dirac deneneracy line in Fig.\ref{Fig:DiracLine} ({\it right}).
These two branches form the Dirac cone, which is on the border between the tilted and overtilted cones.
In other words,  the Hamiltonian (\ref{eq:spectrum2}) describes the bosonic Dirac system, which is on the border between the type-I and type-II.
 }
 \end{figure}

For  $\lambda=\pi/2$, the two branches do not interact with each other and may cross each other, see Fig. \ref{Fig:Spectrum}.  In the mode with $\omega=\gamma H$, the transverse component $M_\perp$ oscillates. This mode is  excited in transverse NMR experiments. The mode with $\omega= \Omega_P$ and with oscillating $M_\parallel$ is excited in longitudinal NMR experiments. In the other language  these two branches correspond  respectively to the optical magnon and the light Higgs mode.\cite{Zavjalov2016,VolovikZubkov2015} The modes do not interact with each other only at $\lambda=\pi/2$ and at $\lambda=0$. Otherwise, these modes interact producing the observed parametric decay of Bose-Einstein condensate of optical magnons to light Higgs modes,\cite{Zavjalov2016} and the repulsion of the levels -- the observed avoiding crossing.\cite{Dmitriev2016}.

 At  $\lambda=\pi/2$ and $\gamma H= \Omega_P$ these two branches cross each other. This is the degeneracy point of the level crossing -- the Dirac diabolical point  in the space of the two parameters, $\gamma H= \Omega_P$ and  $\lambda=\pi/2$.  If one takes into account all three components of magnetic field ${\bf H}$, one obtains the Dirac line (circle)  $H_z=0$,  $\gamma H= \Omega_P$  in the 3D space of magnetic field $(H_x,H_y,H_z)$ in Fig.\ref{Fig:DiracLine} ({\it right}), where the spectrum is degenerate. Close to the Dirac line, the Hamiltonian in Eq.(\ref{eq:spectrum2}) transforms to:
\begin{equation}
\label{eq:spectrumApprox}
{\cal H}({\bf H})-\Omega_P^2\approx  
 \Omega_P (\gamma H - \Omega_P) + \Omega_P (\gamma H - \Omega_P) \tau^3
 -   \Omega_P^2\hat{\bf m}  \cdot \hat{\bf h} \, \tau^1,
\end{equation}
where $\hat{\bf h}={\bf H}/H$. Equation (\ref{eq:spectrumApprox}) is analogous to Eq.(\ref{HamiltonianPolar}), with $\gamma \Omega_P$ and $\Omega_P/\gamma$  playing the roles of Fermi velocity and Fermi momentum, and $\Omega_P^2$ being the analog of gap amplitude.
Since the analog of the Fermi velocity coincides with the derivative of the first term in the right-hand side with respect to $H$, the Hamiltonians (\ref{eq:spectrum2}) and (\ref{eq:spectrumApprox}) describe the bosonic Dirac system, which is on the border between the type-I and type-II.\cite{VolovikZubkov2014,Soluyanov2015,Autes2016,VolovikKuang2017} 
The Higgs mode in Fig. \ref{Fig:Spectrum} is "dispersionless", $d\omega/dH=0$, which is on the border between the tilted ($d\omega/dH<0$) and the overtilted ($d\omega/dH>0$) Dirac cones.

In both cases of fermionic and bosonic spectrum in Fig. \ref{Fig:DiracLine}, the Dirac nodal line has nontrivial topological charge $N_2=1$, see e.g. \cite{Heikkila2011,SatoAndo2017}
\begin{equation}
\label{eq:N2}
N_2=\frac{1}{4\pi i}{\rm Tr}\oint_C dl \,\tau_2\tilde{\cal H}^{-1}\partial_l \tilde{\cal H}
\,.
\end{equation}
Here  $\tilde{\cal H}$ is the traceless part of the matrix ${\cal H}$, and the integral is along the loop $C$ in momentum or parameter space enclosing the Dirac line.
The nontrivial topology means that when the momentum ${\bf p}$ in Fig.\ref{Fig:DiracLine} ({\it left}) or magnetic field ${\bf H}$ in Fig.\ref{Fig:DiracLine} ({\it right}) adiabatically evolves along this loop, the corresponding geometric Berry  phase $\Phi$ changes by $\pi$.

In conclusion, there are two topologically protected Dirac lines in the polar phase of superfluid $^3$He. One of them is fermionic, which lives in the 3D momentum space $(p_x,p_y,p_z)$. It gives rise to the peculiar type of  the effective quantum electrodynamics.\cite{NissinenVolovik2018}   The other one is bosonic and lives in the 3D parameter space $(H_x,H_y,H_z)$. The NMR spectrum near this Dirac line has been experimentally studied in Ref. \onlinecite{Dmitriev2016}.
The next task should be to combine the effects of the two Dirac lines, which form  the 2D degeneracy manifold in the extended 6D momentum+parameter space $(p_x,p_y,p_z,H_x,H_y,H_z)$.  This will involve the effects related to dynamics of Bogoliubov quasiparticles near the fermionic Dirac line interacting with the spin waves in vicinity of the bosonic Dirac line, such as adiabatic Thouless pumping.\cite{Thouless1983,Ryu2016}

I thank Tero Heikkil\"a for interesting discussions, which resulted in this paper. This work has been supported by the European Research Council (ERC) under the European Union's Horizon 2020 research and innovation programme (Grant Agreement No. 694248).


\end{document}